%% file: usenix.tex
\documentclass[letterpaper,twocolumn,10pt]{article}
\usepackage{usenix,epsfig,endnotes}

\newcommand{\stitle}[1]{\vspace{1ex} \noindent{{\bf #1}}}
\newcommand{\ours}[0]{Echo\xspace}

\begin{document}

\date{}

\title{\Large \bf \ours: Efficient Co-Scheduling of Hybrid Online-Offline Tasks for \\
Large Language Model Serving}

\maketitle

\thispagestyle{empty}

\begin{abstract}
    Large language models have been widely deployed in various applications, encompassing both interactive online tasks and batched offline tasks. Given the burstiness and latency sensitivity of online tasks, over-provisioning resources is common practice. This allows for the integration of latency-insensitive offline tasks during periods of low online load, enhancing resource utilization. However, strategically serving online and offline tasks through a preemption mechanism fails to fully leverage the flexibility of offline tasks and suffers from KV cache recomputation and irregular workloads.

    In this paper, we introduce \ours, a collaborative online-offline task serving system, including a scheduler, a KV cache manager, and estimation toolkits. The scheduler and KV cache manager work tightly to maximize the throughput of offline tasks, while the estimator further predicts execution time to ensure online task SLOs. The scheduler leverages the batch information of last iteration to reduce the search space for finding the optimal schedule. The KV cache manager sets the priority of the KV cache based on the type of tasks and the opportunity of prefix sharing to reduce the recomputation. Finally, the estimation toolkits predict the execution time, future memory consumption, and the throughput of offline tasks to guide the scheduler, KV cache manager, and the system deployer.
    Evaluation based on real-world workloads demonstrates that \ours~can increase offline task throughput by up to $3.3\times$, while satisfying online task SLOs.
\end{abstract}

\input{1_introduction.tex}
\input{2_background.tex}
\input{3_overview.tex}
\input{4_scheduler.tex}
\input{5_estimator.tex}

\input{6_implementation.tex}
\input{7_evalutaion.tex}

\input{8_related_works.tex}

{\footnotesize \bibliographystyle{acm}
    \bibliography{sample}}

\end{document}

%% file: 1_introduction.tex
\section[Introduction]{Introduction}
\label{sec:introduction}

Recently, large language models (LLMs)~\cite{openai2024chatgpt,deepseekai2024deepseekv3technicalreport,qwen2025qwen25technicalreport,google2024gemini,touvron2023llamaopenefficientfoundation,glm2024chatglm,abdin2024phi4technicalreport,wake2024yilightningtechnicalreport,xai2024grok,sun2024hunyuanlargeopensourcemoemodel} have achieved remarkable performance in various tasks, such as question answering~\cite{kamalloo2023evaluating,liu2021makesgoodincontextexamples}, planning~\cite{significantgravitas2023autogpt, song2023llmplannerfewshotgroundedplanning} and even solving complex mathematical problems~\cite{openai2024reasonwithllm, qwen2024qwq, wu2024inferencescalinglawsempirical,deepseekai2025deepseekr1incentivizingreasoningcapability}.
Numerous companies have developed their LLMs and provided services, enabling users to interact with these models, such as ChatGPT~\cite{openai2024chatgpt}, Claude~\cite{anthropic2024claude}, DeepSeek~\cite{deepseekai2024deepseekv3technicalreport}, Qwen~\cite{qwen2025qwen25technicalreport}, and Gemini~\cite{google2024gemini}.
However, serving LLMs is costly due to the substantial computational and memory demands, often necessitating expensive accelerators.
Therefore, optimizing resource utilization to achieve economic and efficient serving of LLMs has become a critical challenge for service providers~\cite{chen2024efficienteconomiclargelanguage,miao2024spotserve,gao2024cachedattention}.

In practice, the tasks users submit to LLMs can generally be categorized into two types, online and offline tasks, each with distinct characteristics and requirements.
\begin{itemize}
      \item \textbf{Online tasks}, such as interactive chatbots~\cite{openai2024chatgpt,anthropic2024claude,chiang2024chatbotarenaopenplatform}, programming assistants~\cite{github2024copilot, roziere2023codellama}, real-time translation and recommendation systems~\cite{Hou2023LargeLM,zhao2024recommendersystemseralarge}, are often latency-sensitive and require rapid response time.
            Burstiness~\cite{ali2014measuring, ari2003managing, bodik2010characterizing} and tidal patterns~\cite{benson2010network} of online tasks are common in real-world serving, resulting in resource over-provisioning to meet the peak workload, while over-provisioned resources are underutilized during off-peak periods.
      \item \textbf{Offline tasks}, such as text summarization~\cite{zhang-etal-2024-benchmarking,jin2024comprehensivesurveyprocessorientedautomatic}, data wrangling~\cite{narayan2022foundationmodelswrangledata}, LLM benchmarking~\cite{liang2023holisticevaluationlanguagemodels}, data annotation~\cite{tan2024largelanguagemodelsdata}, and synthetic data generation~\cite{guo2024generativeaisyntheticdata}, have low interactivity and loose latency requirements, and are often committed in a batched manner.
      Service providers often offer a relatively cheaper batched inference API for offline tasks, priced at up to 50\% less than online services, e.g., OpenAI~\cite{openai_batch_api}.
\end{itemize}

Therefore, service providers collaboratively serve both online and offline tasks~\cite{openai_batch_api,claude_batch_api,zhipu_batch_api}, to optimize resource utilization by executing offline tasks during periods of low online task workload.
Extensive research has been conducted on 1) maximizing the throughput of online tasks under SLO requirements with given resources~\cite{liu2024andes,agrawal2024taming,zhong2024distserve}, and 2) maximizing the throughput of offline tasks~\cite{sheng2023flexgenhighthroughputgenerativeinference, cao2024moelightninghighthroughputmoeinference,jin2023s3}.
However, coordinating online and offline tasks has distinct objectives, in this paper, our goal is to \emph{maximize the throughput of offline tasks while ensuring the SLOs of online tasks}.

Intuitively, there are two straightforward strategies to coordinate online and offline tasks: 1) Equally scheduling online and offline tasks, and submitting offline tasks at low workload periods (e.g., night time).
This strategy can improve resource utilization in a certain extent, but it may lead to violations of online SLOs if too many offline tasks are submitted and resources are wasted when the workload is neither high nor low.
2) Assigning higher priority to online tasks and allowing them to preempt offline tasks. This method can ensure high resource utilization but misses the opportunities of leveraging the prior knowledge of offline tasks to improve execution efficiency and reduce the recomputation of the KV cache.
Specifically, frequent preemption of offline tasks with KV cache in memory can lead to significant recomputation. Moreover, offline tasks often have a high prefix sharing rate (e.g., more than 80\%~\cite{srivatsa2024prebleefficientdistributedprompt}), which can be further leveraged but this strategy overlooks it. Finally, the irregular prompt lengths can degrade the system performance~\cite{agrawal2023sarathiefficientllminference}, which must be considered in the scheduling strategy.

To address these weaknesses, we aim to unify the scheduling and KV cache management to maximize the execution efficiency and KV cache reuse, while ensuring the SLOs of online tasks. To this end, we propose \ours, a collaborative online-offline task serving system, which includes a task scheduler, a KV cache manager, and a resource estimator (Section~\ref{sec:overview}).
However, it is challenging to design a collaborative online-offline task scheduling system for LLM serving due to the following reasons:

\begin{itemize}
      \item \textbf{Complexity in batch construction.} The number of possible batch combinations is exponential to the number of requests, which is massive in the offline request pool. Moreover, the two stages of LLM inference and the further chunked prefill strategy make batch construction more complex.
      \item \textbf{KV cache from various tasks with various requirements.} The KV cache is crucial for the efficiency of LLM inference, and the reuse of the KV cache can significantly reduce the recomputation.
            However, the KV cache sourcing from latency-sensitive online tasks and cost-effective offline tasks makes cache management more challenging. The upcoming online task will preempt the offline task, resulting in the recomputation of the offline's KV cache.
            Additionally, the KV cache has different levels of reuse opportunities further complicates the cache management.
      \item \textbf{Interference between KV cache and batch construction.} The KV cache management and batch construction are closely related. A larger batch size can have higher execution efficiency but requires more memory for the KV cache. In contrast, the KV cache in memory also affects the batch construction strategy, as the requests can be scheduled based on the KV cache availability.
      \item \textbf{Uncertainty in execution time and future workload.} The exact execution time and memory usage of the system are unknown before the execution. To ensure the SLOs of online tasks, the scheduler needs to estimate the execution time for the current iteration. To avoid frequent KV cache eviction, the KV cache manager needs to predict the memory usage for the upcoming period. Furthermore, it will be helpful for the system deployers to know the minimal resources required to meet the SLOs of online tasks and the maximum throughput of offline tasks given the available resources.
\end{itemize}

In Section~\ref{sec:cook_scheduler_design}, we delve into the design of the task scheduler and the KV cache manager to address the first three challenges. We observe that the scheduler could \emph{leverage the batch information in last iteration} to guide the batch construction in the current iteration, which significantly reduces the search space. Regarding the KV cache management, we notice that the source of the KV cache and the reuse opportunities can be employed to set the priority of the KV cache in memory.
Finally, considering the interference between the scheduler and the KV cache manager, our scheduler can aware the KV cache status, while the KV cache manager also considers the task information. To handle the KV cache eviction resulted from bursty online tasks, the KV cache manager will leave some space for the upcoming online tasks, while the scheduler will consider the KV cache status to guide the batch construction.

In Section~\ref{sec:estimator}, we propose estimation toolkits to address the last challenge. We first review the execution of a batch in typical LLM inference, and model the execution time of a batch, which is related to the average and maximum length of the requests in the batch. Then we predict the memory consumption of the KV cache for the upcoming period according to the historical traces of online tasks to handle the burstiness of online tasks. Finally, we simulate the scheduler and cache manager to estimate the minimal resources required to meet the SLOs of online tasks and the maximum throughput of offline tasks given the available resources, which helps the system deployers configure the system's resources.

In Section~\ref{sec:evaluation}, we evaluate \ours using real-world traces and datasets. Compared with the baseline strategy, \ours can improve the throughput of offline tasks by up to $3.3\times$, while not violating online SLOs. The further experimental results reveal that the superior performance of \ours is attributed to the better KV cache reuse, which improves the cache hit rate to $78.6\%$ with LooGLE QA\_Short.

%% file: 2_background.tex
\section{Background}
\label{sec:background}

In this section, we introduce the background of LLM inference and the characteristics of LLM serving workloads. We then summarize several insights derived from our observations of these workloads, which motivate the design of our system.

\subsection{LLM Inference}
\label{sec:llm_inference}

\begin{figure}
    \centering
    \includegraphics[width=0.95\columnwidth]{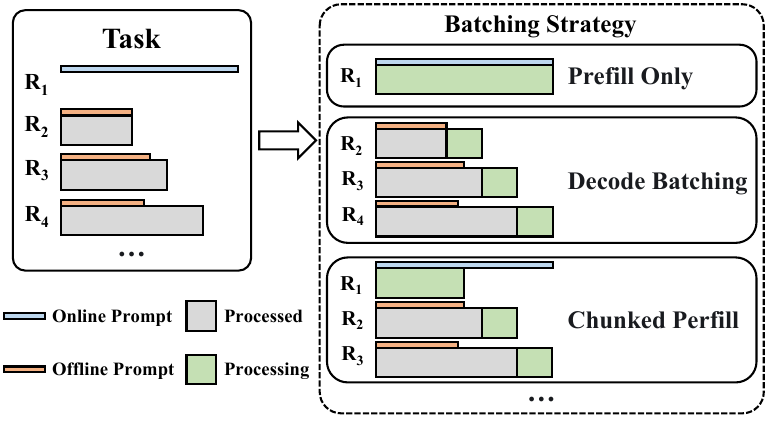}
    \caption{Batching strategy.
        \vspace{-.15in}
    }
    \label{fig:batch strategy}
\end{figure}

\stitle{Prefill and decode.}
The autoregressive inference process of mainstream LLMs can be divided into two stages: prefill and decode.
As shown in Figure~\ref{fig:batch strategy}, the prefill stage processes the user input prompt, generates and caches the key-value cache~\cite{kwon2023efficient, reiner2023scalingtransformer}, which involves a high computational demand and is compute-bound.
The decode stage uses the KV cache to generate a single token at a time, appending the corresponding KV state to the cache.
Compared to the prefill stage, the decode stage has lower computational demand but is memory-bound due to the high frequency of KV cache access.

Many optimization techniques for LLM inference have been proposed to improve the execution efficiency. Typical approaches include:
\begin{itemize}
    \item \textbf{Decode batching}~\cite{yu2022orca,kwon2023efficient,zheng2024batchllmoptimizinglargebatched}. Recongizing the low workload of single request's decode stage, researchers consider batching multiple requests' decode processes together to fully utilize the high parallelism of GPUs.
    \item \textbf{Chunked prefills}~\cite{agrawal2024taming,agrawal2023sarathiefficientllminference}. To fully utilize the memory bandwidth and computational resources of GPUs, researchers suggest batching the compute-intensive prefill request with memory-intensive decode requests. Furthermore, the prefill task can be further chunked to trade off between memory and compute resources. Moreover, chunked prefill can also improve the user experience by reducing the time between tokens.
\end{itemize}

Though these techniques have been shown to improve the efficiency of LLM inference, \emph{they actually introduce complexity to task scheduling and resource management.}

As discussed earlier, the decode iteration of LLM inference is memory-bound, and the batch construction strategy plays a crucial role in enhancing the efficiency of the decode stage.
The batching strategy in LLM inference is a complex problem, as it must consider the length of the input prompt, the number of requests in a batch, the distribution of prefill/decode tasks, and the potential reuse of the KV cache.

\subsection{Online Task}

Common LLM services, such as chatbots~\cite{openai2024chatgpt,anthropic2024claude,chiang2024chatbotarenaopenplatform} and programming assistants~\cite{github2024copilot, roziere2023codellama}, are typically deployed as online services, which are highly interactive and require real-time response to the requests from users, with high throughput and low latency as key performance indicators~\cite{liu2024andes}.

\stitle{Service level objectives (SLOs).}
According to~\cite{patel2023splitwise,agrawal2024taming,zhong2024distserve, qin2024mooncakekvcachecentricdisaggregatedarchitecture}, the SLOs of online tasks include time-to-first-token (TTFT) representing the response time to user input, and time-per-output-token (TPOT) representing the output speed.
Considering the chatbot application, after the user sends a message, the chatbot must respond within a specified time (e.g. TTFT=1 s), and the subsequent tokens must be generated at a speed faster than human reading speed (e.g. TPOT=180 ms)~\cite{brysbaert2019many}.
Many researches focus on improve the latency of online tasks. Andes\cite{liu2024andes} first proposed Quality of Experience-aware LLM serving system. Sarathi-Serve~\cite{agrawal2024taming} mitigate generation stalls with chunked prefills. Some recent works~\cite{patel2023splitwise,zhong2024distserve,hu2024inferenceinterferencedisaggregatellm} obeserve the opportunities of disaggregating the prefill and decode stages to reduce the latency of LLM serving.

In the online task-only workload, a simple scheduler follows a First-Come-First-Serve (FCFS) policy, which batches the decode stage of active online requests and schedules the prefill stage of online requests in order.

\begin{figure}[t]
    \centering
    \includegraphics[width=.95\linewidth]{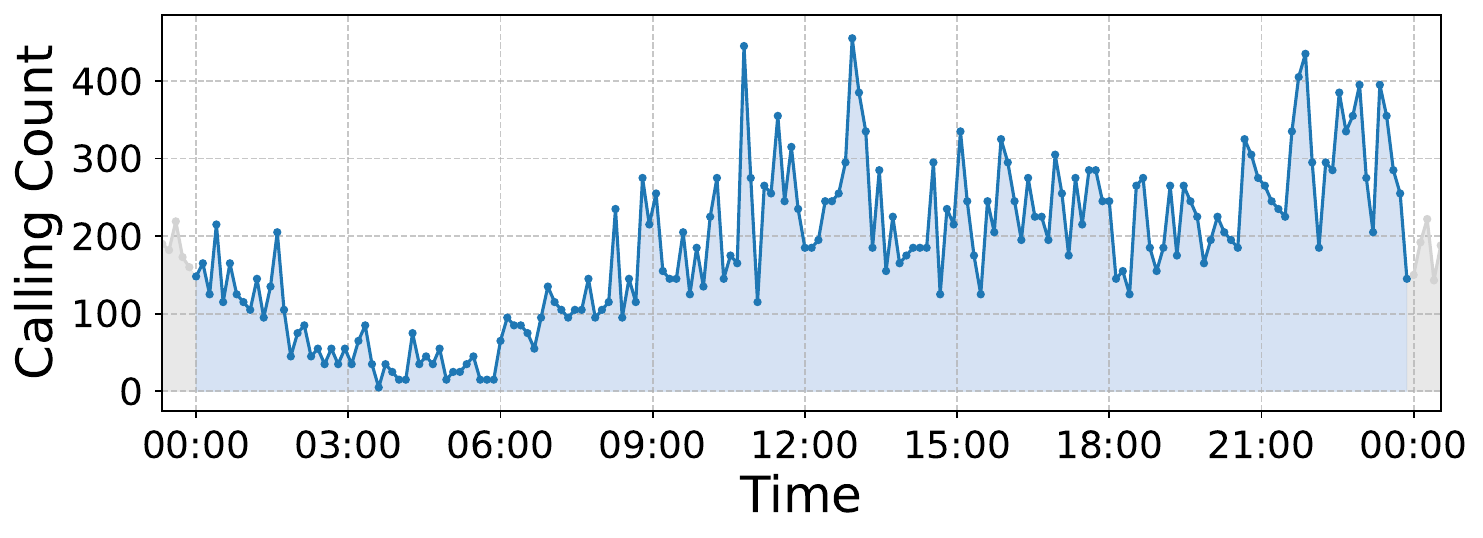}
    \caption{A 24-hour trace of a typical online task.\vspace{-.15in}}
    \label{fig:trace-close}
\end{figure}

\stitle{Burstiness and tidality.}
Figure \ref{fig:trace-close} illustrates the 24-hour call pattern trends from a real-world trace by an LLM service provider. We observe that 1) on short time scales, the burstiness of online tasks typically occurs, such as around 13:00, and 2) on longer time scales, the workload follows a tidal pattern, with peak usage periods during 12:00-14:00 and low usage periods during 04:00-06:00, where the request volume during peak hours is approximately $6 \times$ higher than that during off-peak hours.

To handle the burstiness and ensure the SLOs of online tasks, \emph{service providers often overprovision resources}~\cite{qiao2024conserveharvestinggpuslowlatency}, leading to low resource utilization and high costs.

\subsection{Offline Task}

In contrast to online tasks, offline tasks are latency-insensitive and can be submitted in a batched manner, typical applications include text summarization~\cite{zhang-etal-2024-benchmarking,jin2024comprehensivesurveyprocessorientedautomatic}, long-document QA~\cite{li2024looglelongcontextlanguagemodels}, data wrangling~\cite{narayan2022foundationmodelswrangledata}, LLM benchmarking~\cite{liang2023holisticevaluationlanguagemodels}, data annotation~\cite{tan2024largelanguagemodelsdata} and synthetic data generation~\cite{guo2024generativeaisyntheticdata}.
The loose constraints allow more flexibility in scheduling and resource utilization for service providers to optimize serving offline tasks, which is reflected by cheaper API costs for offline tasks~\cite{openai_batch_api}.

\begin{table}[ht]
    \centering
    \caption{Prefix sharing rate of different workloads~\cite{srivatsa2024prebleefficientdistributedprompt}.}
    \resizebox{\linewidth}{!}{
        \begin{tabular}{cccc}
            \toprule
            Mode                     & Workload                                                            & Avg. Prompt & Shared Rate \\
            \midrule
            Online                   & ShareGPT~\cite{eccleston2023sharegpt}                               & 308         & < 5\%       \\
            \hline
            \multirow{3}{*}{Offline} & LooGLE~\cite{li2024looglelongcontextlanguagemodels}                 & 23,474      & 91\%        \\
                                     & ToolBench~\cite{guo2024stabletoolbenchstablelargescalebenchmarking} & 1,835       & 85\%        \\
                                     & NExT-QA~\cite{xiao2021nextqanextphasequestionansweringexplaining}   & 9,865       & 88\%        \\
            \bottomrule
        \end{tabular}
    }
    \label{tab:prefix_sharing_modified}

\end{table}

As shown in Table~\ref{tab:prefix_sharing_modified}, compared to online tasks, offline tasks have following characteristics:

\stitle{Higher prefix-sharing opportunities.}
Prefix caching have been widely used in LLM inference frameworks~\cite{kwon2023efficient, zheng2024sglangefficientexecutionstructured, qin2024mooncakekvcachecentricdisaggregatedarchitecture} to reduce the recomputation of prefix tokens.
Offline tasks present valuable opportunities for reorganizing the cache to enhance the reuse of prefix tokens with spatial locality. This approach can substantially reduce recomputation and improve the efficiency of the prefill stage. Fortunately, as shown in Table~\ref{tab:prefix_sharing_modified}, batched long requests usually have a high prefix-sharing rate, which can be leveraged to improve the efficiency of offline tasks.

\stitle{Long offline prompts.}
As shown in Table~\ref{tab:prefix_sharing_modified}, due to the inherent characteristics of offline tasks, the prompt length of offline tasks is typically longer than that of online tasks. For instance, QA tasks, whose prompts are usually long articles or books, have an average prompt length of 9,865 tokens, and is a typical example of offline tasks.

%% file: 3_overview.tex
\section{Overview}
\label{sec:overview}

In this section, we begin by providing an overview of modeling the online-offline task scheduling problem in the context of LLM serving. This overview aims to clarify the requirements and challenges of the system.
Following this, we will outline the architecture of \ours, which comprises three key components: a task scheduler, a KV cache manager, and a resource estimator.

\subsection{Model Formulation and Challenges}
\label{subsec:modeling_online_offline_scheduling}

\stitle{Benefit of executing an iteration.}
As indicated in Section~\ref{sec:llm_inference}, there exist various combinations of prefill and decode tasks in a batch of requests executed in an iteration, while various batch configurations may lead to different execution efficiency.
Many researches~\cite{kwon2023efficient,zhong2024distserve,cao2024moelightninghighthroughputmoeinference} use throughput as the primary metric to evaluate the end-to-end system performance, which is defined as the number of tokens generated per second. While for batch-oriented scheduling, this end-to-end metric is not suitable, as the throughput of a batch is contibuted by the tokens processed in prefill stage and generated in decode stage.
Therefore, we first define the benefit of executing a batch of requests in an iteration.

Generally, the benefit of a batch can be measured by the tokens processed in the batch. For a request, we denote the benefit of prefilling or decoding the $i$-th token as $b(P_i)$ and $b(D_i)$, respectively. Briefly, we use $T$ to denote the set of tokens in the batch including both prefill and decode tokens, and the benefit of the current iteration can be defined as follows:
\begin{equation}
    \begin{aligned}
        \text{Benefit} = \sum_{t\in T} b(t).
    \end{aligned}
\end{equation}

\emph{Challenge 1: The number of choices for the batch configuration (scheduling decisions) is exponential.}

\stitle{Punishment for evicting the KV cache.}
When GPU memory becomes full, some KV cache entries (mostly from offline requests) must be evicted to make room for new entries generated by ongoing requests (mostly from online requests). However, the evicted KV cache may still be needed in the future for two reasons: (1) the corresponding request has not yet been completed, or (2) a future request may share the same prefix as the evicted KV cache. As a result, evicting KV cache entries can lead to the need for recomputation in the future, incurring a penalty defined as follows:
\begin{equation}
    \label{eq:punishment}
    \begin{aligned}
        \text{Punishment} = \sum_{t\in \Delta C^*} b(t).
    \end{aligned}
\end{equation}
Here, $\Delta C^*$ denotes the set of evicted tokens from the KV cache that will be required in the future, and thus, the punishment reflects the cost of re-prefilling the KV cache for the evicted tokens.

\emph{Challenge 2: The reuse of the KV cache and different properties of online and offline requests complicate the management of the KV cache.}

\stitle{Relationship between scheduled tokens and the KV cache.}
Noticing that the scheduled tokens is constrained by the KV cache, i.e., for a request $r$, its tokens can be scheduled only if the corresponding KV cache before the tokens are in the memory.
This constraint introduces a dependency between the scheduler and the KV cache manager, as the scheduler needs to consider the KV cache status when scheduling requests, while the KV cache manager needs to evict KV cache entries to make room for new entries generated by ongoing requests.

\emph{Challenge 3: The interference between the KV cache and batch construction complicates the system design.}

\stitle{Time and memory constraints.}
For $i$-th iteration, the execution time $\text{Time}_i$ should be within the SLO, and the memory usage $\text{Memory}_i$ should not exceed the GPU memory limit.
However, we can not get the exact value before executing the iteration. Therefore, we need to estimate the execution time and future memory usage, which is non-trivial due to the complex nature of LLM inference and unpredictable workload.

\emph{Challenge 4: The estimation of execution time and memory usage is non-trivial.}

\stitle{Optimization Problem.}
Our objective is to maximize the benefit per second, which can be understood as the system's throughput.
This throughput should be evaluated from a long-term perspective and encompasses several iterations,
where we consider $N$ iterations in total and use the sets $\mathbf{T}=\{T_1,T_2,\ldots,T_N\}$ and $\mathbf{C}=\{C_1,C_2,\ldots,C_N\}$ to denote processed tokens and KV caches, respectively, with the subscript $i$ indicating the $i$-th iteration.
Furthermore, the evicted KV cache set is denoted as $\Delta C_i= C_{i} - C_{i+1}$ and $\Delta C^*_{i}\subseteq \Delta C_i$.

Briefly, we formulate the following optimization problem:
\begin{equation}
    \label{eq:optimization}
    \begin{aligned}
        \underset{\color{blue}\mathbf{T},\mathbf{C}}{\arg\max} \quad & \frac{\sum_{i=1}^{N}\left(\text{Benefit}_i-\text{Punishment}_i\right)}{\sum_{i=1}^{N}\text{Time}_i},          \\
        \text{s.t.} \quad                                            & \forall i, {\color{blue}\text{Time}_i} \leq \text{SLO}, {\color{blue}\text{Memory}_i} \leq \text{GPU Memory}.
        \\
    \end{aligned}
\end{equation}

\subsection{Architecture of \ours}
\begin{figure}[t]
    \centering
    \includegraphics[width=1\linewidth]{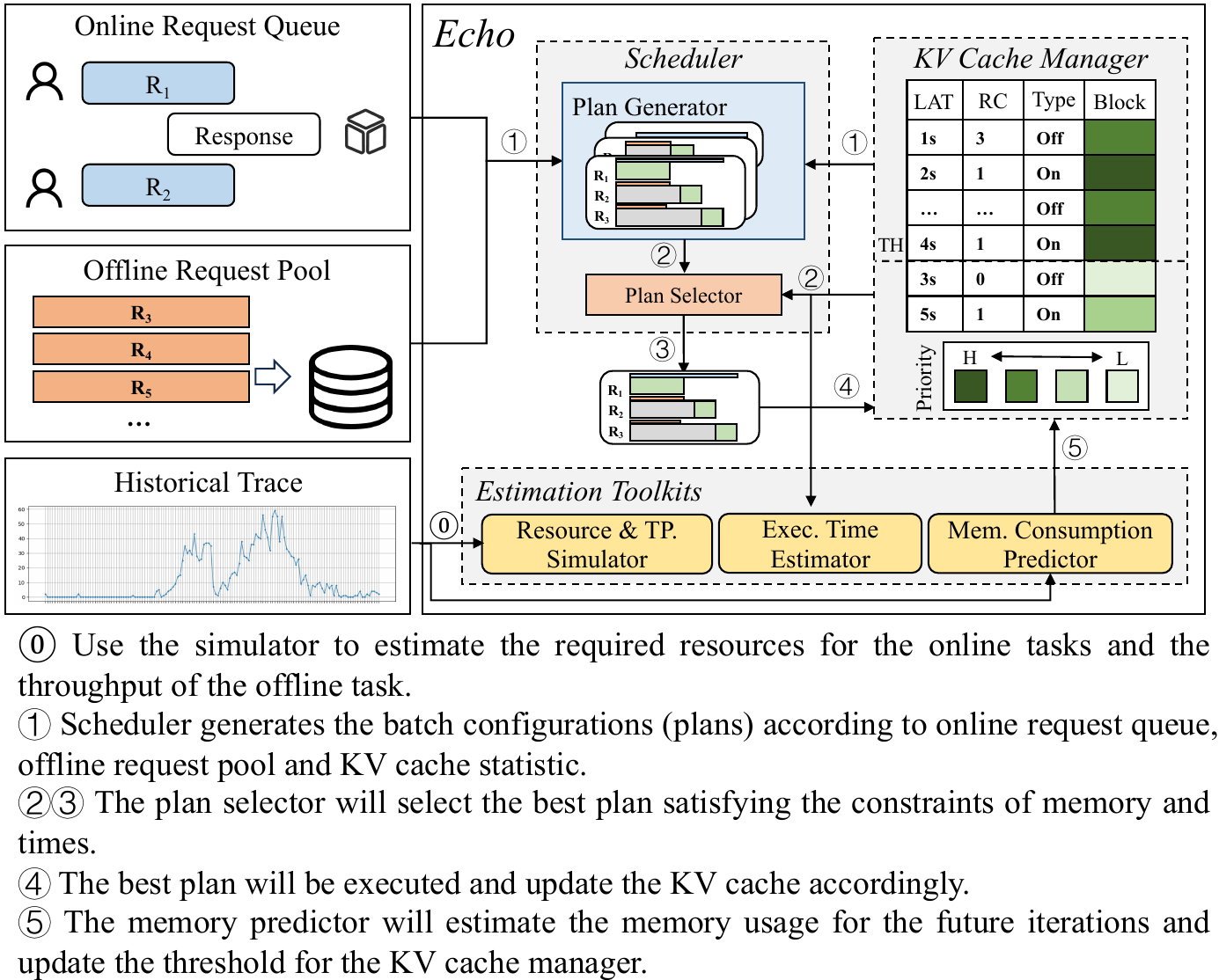}
    \caption{Overview of \ours.\vspace{-.15in}}
    \label{fig:overview}
\end{figure}

After reviewing this optimization problem (especially the variables highlighted in blue in Eq.\ref{eq:optimization}), we propose \ours, which consists of three key components:

\begin{itemize}
    \item \textbf{KV cache-aware task scheduler.} The scheduler determines the request and executed tokens in current iteration. It includes the plan generator and plan selector.
          The plan generator will generate the batch configurations (plans in Figure~\ref{fig:overview}) based on the input requests and KV cache status. The plan selector will select the best plan (with the highest benefit per second) satisfying the memory and time constraints. Therefore, the plan selector will interact with the KV cache manager and the execution time estimator to make the decision.
          After the plan is selected, the scheduler will submit the plan to the execution engine to execute the batch of requests.
    \item \textbf{Task-aware KV cache manager.} The manager decides the KV cache entries in memory, and when the GPU memory is full, it needs to evict some KV cache entries to make room for new entries.
          In addition to the traditional LAT (last access time) information of the KV cache, we further incorporate the future access information of the KV cache, denoted RC (abbreviation of reference count), to indicate the number of requests that will access the KV cache in the future. Moreover, according to the RC information and the source of the request, i.e., online or offline, the manager will set the priority of the KV cache entries. As shown in Figure~\ref{fig:overview}, the second entry in the KV cache has a lower RC value compared to the first entry, but it is from an online request, so it has a higher priority.
          Moreover, the manager will set the threshold for the KV cache size of running requests, leaving room for bursty online requests. Therefore, the manager will interact with memory consumption predictor to make the decision.
    \item \textbf{Estimation toolkits.} The toolkits include three components: an execution time estimator, a memory consumption predictor, and a resource and throughput simulator. The execution time estimator estimates the execution time of the batch generated by scheduler and return feedbacks for decision. The memory consumption predictor predicts the future memory usage of the system and provides the threshold for KV cache manager. The resource and throughput simulator simulates the resource usage and throughput according to historical data and gives suggestions for the system deployer to adjust the resource allocation.
\end{itemize}

The bottom part of Figure~\ref{fig:overview} further illustrates the workflow of \ours. Briefly, in each iteration, the task scheduler determines the batch configuration according to input requests, KV cache status provided by the KV cache manager, and the estimation of execution time. Subsequently, the batch is executed, and the KV cache manager manages the KV cache entries in GPU memory. Finally, the future memory usage are predicted to guide the KV cache manager to set the threshold for KV cache size of running requests.

%% file: 4_scheduler.tex
\section{Scheduler and KV Cache Manager Design}
\label{sec:cook_scheduler_design}

In this section, we extend the abstract model presented in Section~\ref{subsec:modeling_online_offline_scheduling} to a more concrete model, which clearly describes the challenges of designing \ours.
Subsequently, we delve into the design of the \ours scheduler, starting with several naive strategies and then introducing our proposed strategies. Finally, we present the design of the KV cache manager.

\subsection{Scheduling Strategies}

As discussed in Section~\ref{sec:introduction}, treating offline tasks as online tasks will result in SLO violations or resource overprovisioning, thus we omit this strategy.
Our strategy is founded on the preempting-based scheduling policy~\cite{vllm_project_pull_5958}, which allows the scheduler to preempt offline tasks to accommodate online tasks.

However, for each iteration, the combination of batch is up to exponential. Particularly, for each request, we need to decide whether to schedule it, resulting in a combinatorial optimization problem with $O(2^N)$ complexity, where $N$ is the number of requests, which can be extremely large (LooGLE~\cite{li2024looglelongcontextlanguagemodels}, a dataset for document-based QA, consists of 1,951 requests) when considering the offline request pool.

\stitle{Problem Simplification.}
Actually, the vast decision space of the scheduling problem can be simplified by considering the following observations:
\begin{center}
    \framebox{
        \begin{minipage}{0.9\linewidth}
            \centering
            \emph{Last batch can help to determine the current batch.}
        \end{minipage}
    }
\end{center}
Assuming last batch is optimal, current batch refers it and only conducts minor adjustments, including 1) evicting a offline request to accommodate a online request or fit the tight SLOs, 2) scheduling a prefill task (either online or offline) for further decoding, and 3) scheduling a offline decode task whose KV cache is in the GPU memory.

Noticing for online tasks, they are always scheduled in the FCFS (First-Come-First-Serve) order to ensure the SLOs. Therefore, we focus on the scheduling of offline tasks.

\begin{figure}[htbp]
    \centering
    \includegraphics[width=1\linewidth]{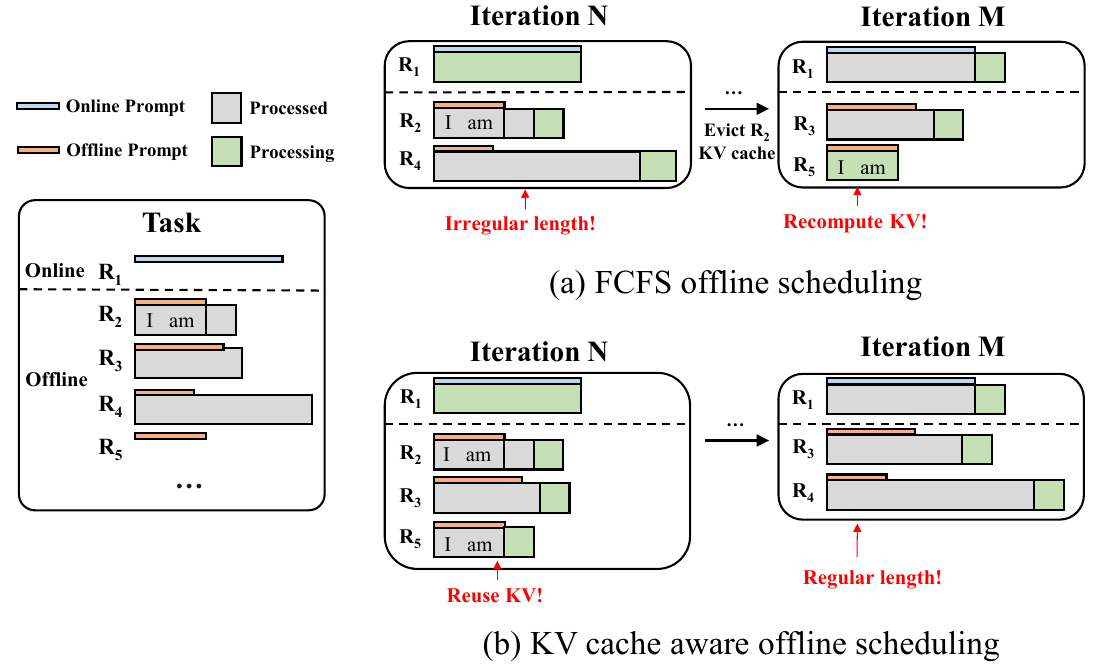}
    \vspace{-.15in}
    \caption{Different strategies.\vspace{-.15in}}
    \label{fig:sub2}
\end{figure}

\stitle{FCFS offline scheduling.}
The scheduler add offline requests to the batch in the FCFS order, which is a straightforward strategy.
As shown in Figure~\ref{fig:sub2}(a), it schedules offline requests from the offline request pool following the top-down order. Obviously, this strategy suffers from following issues:
\begin{itemize}
    \item \textbf{KV cache recomputation for same prefix.} We notice that Requests $R_2$ and $R_5$ share the same prefix ``I am'', which means that the KV cache of $R_2$ can be reused by $R_5$. However, since $R_2$ and $R_5$ are scheduled in different iterations, the KV cache of $R_2$ has been evicted when $R_5$ is scheduled, resulting in recomputation of the same prefix.
    \item \textbf{Low execution efficiency due to irregular batching.} Since the scheduler follows a given order to schedule offline and online tasks, a single batch may contain a mix of short online requests ($R_2$) and long offline requests ($R_4$) (or other irregular combinations), which results in low execution efficiency.
\end{itemize}

\stitle{KV cache aware offline scheduling.}
Recalling the optimization problem, a natural strategy is to maximize the reward of the current iteration. Specifically, for iteration $i$, we have the following optimization objective:
\begin{equation}
    \label{eq:optimization single iteration}
    \begin{aligned}
        \underset{T_i,C_i}{\text{maximize}} \quad & \frac{\text{Benefit}_i-\text{Punishment}_i}{\text{Time}_i}
    \end{aligned}
\end{equation}

Instead of launching or preempting offline requests according to the FCFS policy, we try all offline requests in the offline request pool and select the best one to add to the batch or all offline requests in the batch and select the best one to preempt. We choose the best one based on the estimated execution time which will be discussed in Section~\ref{subsec:batch_execution_time_estimation} and the reward of the current iteration defined in Eq.~\ref{eq:optimization single iteration}.

As shown in Figure~\ref{fig:sub2}(b), the scheduler selects the offline requests $R_2$ and $R_5$ with same prefix into the batch of iteration $n$ to maximize the KV cache reuse. For requests $R_4$, it will be scheduled when the requests in the batch having long KV cache, which improve the regularity of the batch.

However, if we rely solely on the scheduler, the system will trap into a local optimal, as the scheduler will prefer scheduling more offline requests to maximize throughput~\cite{he2024fastdecode,sheng2023flexgenhighthroughputgenerativeinference,xu2024pie}.
\emph{This results in full occupation of GPU memory and frequent KV cache eviction of running offline requests.}

\subsection{Task-aware KV Cache Manager Design}
\label{subsec:kv_vache_manager_design}

\begin{figure*}[htbp]
    \centering
    \includegraphics[width=1\linewidth]{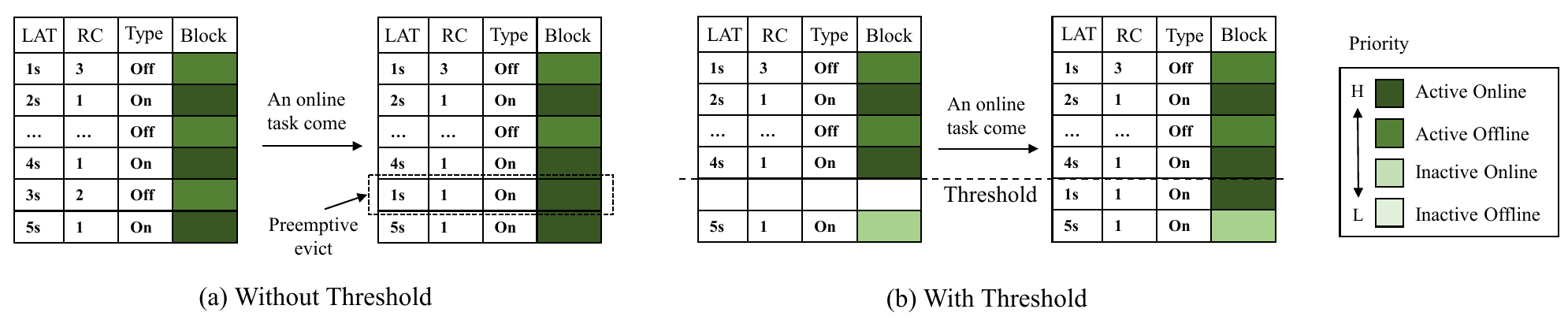}
    \vspace{-.2in}
    \caption{Impact of threshold on KV cache management.\vspace{-.15in}}
    \label{fig:sub1}
\end{figure*}

Considering the traditional LRU (Least Recently Used) policy for KV cache management~\cite{valentin2019lru,kwon2023efficient}, a new incoming online request will evict the KV cache of the least recently used request, which may be an offline request that will be rescheduled in the future. This results in the need for recomputation of the KV cache, incurring a penalty. Accordingly, our key idea is
\begin{center}
    \framebox{
        \begin{minipage}{0.9\linewidth}
            \centering
            \emph{Leave sufficient space for bursty online tasks.}
        \end{minipage}
    }
\end{center}
In other words, we need to limit the KV cache size for active requests, and thus avoid KV cache eviction of running offline requests when new online requests arrive.

\stitle{Priority-based KV cache eviction.}
To identify the useful KV cache entries for future reuse, we can add reference counts to each block of the KV cache (the KV cache is devided into fixed-sized blocks~\cite{kwon2023efficient}, enabling efficient and non-contiguous memory allocation), indicating how may offline requests (including current running request) will reuse it in the future. In addition, we observe different priorities should be set for latency-sensitive online tasks and latency-insensitive offline tasks.
The following is the priority list for KV cache eviction, from highest to lowest:
\begin{itemize}
    \item \textbf{Tokens of running online tasks (priority=$\infty$).} The KV cache of running online tasks should be prioritized for retention, as these tasks are latency-sensitive and will be scheduled in the next iteration.
    \item \textbf{Tokens of active offline tasks with reference counts $rc>0$ (priority=$rc$).} The KV cache of offline tasks with high reference counts should be retained first, as they will be (1) reused in the future (evicting them will result in punishment), and (2) provide more flexibility to the scheduler by allowing scheduling of multiple offline requests requiring these KV cache entries.
    \item \textbf{Tokens of finished online tasks (priority=0.5).} The KV cache of finished online tasks might be reused in future online tasks. Thus, if the GPU memory is not fully occupied, we can keep them for potential future reuse.
    \item \textbf{Tokens of finished offline tasks with no reference counts, i.e., $rc=0$ (priority=0).} The KV cache of offline tasks with no reference counts can be evicted first, as they will not be reused in the future\footnote{Actually, future online tasks may reuse them, but the probability is low and can be ignored.}.
\end{itemize}

As shown in Figure~\ref{fig:sub1}, in addition to the block of KV cache (Fourth column), we also record three additional metadata columns for management, from left to right: the last access time (LAT), the reference count (RC), and the type of the task of the KV cache entry. The last access time is used to determine the least recently used KV cache entry, while the reference count and the type of the task are used to determine the priority of the KV cache entry for eviction.
When evicting the KV cache, we will first consider the priority of the KV cache entry, and then the last access time. If the priority is the same, we will evict the least recently used KV cache entry. This strategy prioritizes the retention of KV cache entries that will be reused in the future and source from latency-sensitive online tasks.

\stitle{Threshold to limit the KV cache size for active requests.}
We will set a threshold to limit the KV cache size for running online tasks and active offline tasks while leaving sufficient space for future bursty online tasks.
The threshold can be determined by predicting the memory consumption of future bursty online tasks, as discussed in Section~\ref{subsec:memory_consumption_of_kv_cache_estimation}.
On the other hand, this threshold will impact the descision of the scheduler by adjusting the memory contraints in Eq.~\ref{eq:optimization}.

As shown in Figure~\ref{fig:sub1}(a), without the threshold, the GPU memory is fully occupied by the KV cache of running online tasks and active offline tasks, leading to KV cache eviction of offline tasks when new online requests arrive. This actually results in recomputation of the KV cache of these offline tasks, incurring a penalty. By setting the threshold, we can leave sufficient space for bursty online tasks. As depicted in Figure~\ref{fig:sub1}(b), the GPU memory is not fully occupied by running online tasks and active offline tasks, and no useful KV cache is evicted when new online requests arrive.

%% file: 5_estimator.tex
\section{\ours Estimator Design}
\label{sec:estimator}

Similarly, we first extend the constraints in Eq.~\ref{eq:optimization} to a more concrete form, which motivates the design of the \ours estimator.
Subsequently, we present the design of the \ours estimator.

\subsection{Concrete Constraints and Responsibilities}

\stitle{Constraints of online task SLOs.}
As indicated in~\cite{patel2023splitwise,agrawal2024taming,zhong2024distserve,qin2024mooncakekvcachecentricdisaggregatedarchitecture}, the SLO of an online task is defined by two latency constraints: TTFT and TPOT.
We follow~\cite{agrawal2024etalonholisticperformanceevaluation,wang2024revisitingslogoodputmetrics}, which defines the latency function for each output token. For $i$-th token, the latency is defined as: $\text{Latency}_i = \text{TTFT} + i \cdot \text{TPOT}$.
Accordingly, the SLO of the request $r$ in current iteration is defined as: $\text{SLO}_r = \text{Latency}_i-\text{WaitingTime}$, where $i$ is the index of the token output in the current iteration, and $\text{WaitingTime}$ is the time interval between the request arrival and the current iteration.
Thr SLO of the batch of requests in the current iteration is determined by the tightest constraint among all requests in the batch, i.e., $\text{SLO} = \min(\text{SLO}_r)$.

\stitle{Constraints of KV Cache.} The usage of the KV cache should not exceed the maximum memory size. Briefly, we can derive the maximum number of tokens whose KV cache can be kept in the GPU memory, denoted as $N_{\text{KV}}$. Hence, the constraint can be formulated as follows:
\begin{equation}
    |C\cup T| \leq N_{\text{KV}}
\end{equation}

According to the constraints as well as the design of the scheduler and KV cache manager, we summarize the responsibilities of the \ours estimator as follows:

\begin{enumerate}
    \item Estimating the execution time of a batch of requests, which ensures that the online tasks meet their SLOs and guides the scheduler for decision-making by estimating the benefit per unit of time of executing a given batch.
    \item Predicting the memory consumption of the KV cache for online requests in a future time window. Notice that the memory consumption of current iterations can be exactly calculated, while the KV cache manager needs to handle the bursty workload of online tasks to avoid frequent evictions and recomputations.
    \item Suggesting the allocation of resources. We note that with limited resources, the system may not be able to meet the SLOs of online tasks even no offline tasks are executed. Therefore, the \ours estimator should provide guidance on the minimal resources required to meet the SLOs of online tasks. Moreover, the \ours estimator should also estimate the maximum throughput of offline tasks given the available resources. Both estimations are crucial for system deployers to configure the system according to their workload.
\end{enumerate}

\subsection{Batch Execution Time Estimation}
\label{subsec:batch_execution_time_estimation}

As there are various batching configurations, the estimation is non-trivial. Moreover, the combination of prefill and decode requests in a batch further complicates the estimation. We will start by modeling the execution time of prefill-only and decode-only batches, and then extend the model to batches with both prefill and decode requests.

\stitle{One prefill batch.}
It is common practice to process prefill requests one by one~\cite{agrawal2024taming,zhong2024distserve}. In the prefill stage, the cost of the attention layers is quadratic to the sequence length $l$, while the cost of the other layers is linear to $l$. Therefore, we model the prefill time as follows:
\begin{equation}
    \text{Time}_\text{prefill} = \max(\alpha l^2 + \beta l, c)
\end{equation}
where $\alpha$ and $\beta$ are the coefficients of the quadratic and linear terms, respectively, and $c$ is a constant term that reflects the minimum time needed to prefill a request, which may not fully utilize the availabel resources.

\stitle{Decode-only batch.}
For decode requests, batch processing can improve resource utilization. Consider a batch of requests $R=\{r_1, r_2, \ldots, r_n\}$, where $r_i$ has a sequence length of $l_i$. We denote the sequence lengths as $L=\{l_1, l_2, \ldots, l_n\}$.
In fact, the execution time of a batch of requests can be modeled as a pooling operation as follows:
\begin{equation}
    \text{Time}_\text{decode} = \gamma \max(L) + \delta \text{mean}(L),
\end{equation}
where $\gamma$ and $\delta$ are the coefficients of the max and mean pooling operations, respectively.

\stitle{Batch with prefill and decode.}
Previous work\cite{agrawal2024taming} argues that batching prefill and decode requests together can improve the resource utilization, as prefill requests are compute-bound and decode requests are memory-bound. Therefore, the execution time of such a batch is greater than the maximum of the prefill and decode times but less than their sum. This relationship can be modeled as follows:
\begin{equation}
    \begin{aligned}
        \text{Time}_\text{batch} = & \lambda \max(\text{Time}_\text{prefill}, \text{Time}_\text{decode})        \\
                                   & + (1-\lambda) \min(\text{Time}_\text{prefill}, \text{Time}_\text{decode}),
    \end{aligned}
\end{equation}
where $\lambda$ is the coefficient that balances the two terms.

To determine the values for $\alpha$, $\beta$, $\gamma$, $\delta$, and $\lambda$, we conduct micro-benchmarks before deploying the system.

\subsection{Memory Consumption of KV Cache Prediction}
\label{subsec:memory_consumption_of_kv_cache_estimation}

Predicting the memory consumption of the KV cache for future requests is challenging, as the workload of online tasks varies in the long term (e.g., day and night) and is bursty in the short term (e.g., a sudden surge of requests within a minute).
Fortunately, we observe that, \emph{the extent of the burstiness (i.e., variance) and the average workload in the medium term (e.g., an hour) remain relatively stable}.

Accordingly, we leverage historical traces of online tasks over the past hour to estimate the possible memory consumption of the KV cache in the next few minutes. Specifically, we calculate the average memory consumption $\mu$ and the variance of memory consumption $\sigma^2$ over the past hour.
Assuming the memory consumption of the KV cache follows a normal distribution, we set the memory consumption of the KV cache to $\mu + 2\sigma$ to ensure the system can effectively handle 95\% of cases.\footnote{2 is a hyperparameter that can be tuned.}

\subsection{Simulation for Resource and Throughput Estimation}
\label{subsec:simulation_of_resource_and_throughput_estimation}
With the estimators of execution time and memory consumption, and by incorporating the historical traces of online and offline tasks,
we can simulate the scheduler and cache manager to estimate the minimual resources required to meet the SLOs of online tasks and the maximum throughput of offline tasks given the availabel resources.

To be specific, the simulation will include the following steps:

\stitle{Step 1: Resource estimation in peak workload.} To reduce simulation complexity, we focus on a short period of time (e.g., 5 minutes) during the peak workload of online tasks. We then enumerate the resources from the smallest to the largest until the online task SLOs are met.

This estimation will be reported to system deployers, who can configure the system's resources accordingly to check whether the it can meet the throughput requirements of offline tasks.

\stitle{Step 2: Throughput estimation of offline tasks.} With the given resources, we can simulate the scheduler and cache manager over an extended period of time (e.g., 1 day) to estimate the maximum throughput of offline tasks.

%% file: 6_implementation.tex
\section{Implementation}
\label{sec:implementation}

We implement \ours and the baselines on the top of vLLM 0.6.3~\cite{kwon2023efficient}. The scheduler and KV cache manager serves as the upper layer of \ours, and the estimator is a separate component at the lower layer that provides black-box estimates for the scheduler.

\stitle{Scheduler.}
To extend vLLM to achieve SLO-awareness, the \ours scheduler schedules online and offline requests by calling the estimator in the upper layer, allowing offline requests to be scheduled in batch without violating the SLOs of online requests.

Specifically, in each iteration, the scheduler first adds requests from the online queue to the batch under the constraints of the KV cache and SLOs, and calls the estimator to predict the benefit of the current batch. Only when all online requests in the queue are added to the batch, the scheduler selects requests from the offline request pool. Similarly, the scheduler selects the appropriate offline requests based on the prediction results of the estimator to maximize the benefit and minimize the punishment. Finally, the scheduler sends the selected requests in the batch to the vLLM engine to execute a step, completing an iteration.
Note that the SLOs and estimator strategies of all SLO-aware baselines are completely consistent to ensure the fairness of the experiments.

\stitle{KV cache manager.} We implement \ours KV cache manager based on the design in Section~\ref{subsec:kv_vache_manager_design}. Based on the Automatic Prefix Caching (APC) design~\cite{vllm_apc}, we extend the LRU strategy of vLLM and design multiple priorities for block eviction based on workload and reference count. In the free table organized based on the priority queue, the block with the lowest priority (as described in Section~\ref{subsec:kv_vache_manager_design}) will be evicted first. Blocks with the same priority will be evicted according to the default LRU strategy of vLLM. This simple design efficiently avoids the frequent flushing of the prefix cache of offline tasks with high sharing rates by online tasks when GPU memory is tight.

\stitle{Estimator.}
We implement the estimator by python and split the estimator into a separate component to provide black-box estimates for the scheduler flexibly according to the characteristics of the workload and the requirements of different SLOs. The estimator is based on hardware-related profile information, and calculates the benefits and costs according to the strategies in Section~\ref{sec:estimator}, serving the upper-layer calls from the scheduler. Before execution, we conduct a series of micro-benchmarks to configure the hyperparameters of the estimator.

\stitle{Online queue and offline pool.}
We organize online requests as a queue, which is naturally suitable for scheduling for FCFS strategy. To construct the offline candidate pool efficiently and ensuring the temporal locality of the prefix, we coarsely divide offline requests into different buckets based on the length distribution, where each bucket contains requests with similar lengths and is organized in a radix tree. In each iteration, the scheduler selects requests from the tree corresponding to the appropriate bucket based on the current batch and the KV cache to add to the batch.

\stitle{Preemption.} We use recomputing mode of vLLM to support preemption, in which the scheduler preempts and release the KV cache of the victim request, and then re-executes the request in the future. The swapping mode of vLLM is out of our consideration, since our optimization is not related to the swapping mechanism.

%% file: 7_evalutaion.tex
\section{Evaluation}
\label{sec:evaluation}
In this section, we evaluate the performance of \ours~on real-world online serving traces with different offline datasets. Through experiments, we aim to answer the following questions:
\begin{itemize}
    \item \emph{The overall performance.} How much throughput speedup does \ours~bring to offline tasks compared to baseline strategies? Can \ours ensure that online tasks meet SLOs?
    \item \emph{Reasone of performance improvement.} What is the reason for the performance improvement of \ours? We will dive into the detailed metrics of the KV cache hit rate and the memory usage of various tasks.
    \item \emph{The accuracy of the estimator.} How accurate is the estimator in predicting the memory usage?
\end{itemize}

\subsection{Experimental Setup}
\stitle{Models and environment.}
We conduct our experiments on a server equipped with a 112-core CPU and 512GB of memory, and an NVIDIA A100-PCIE-40GB GPU, running Debian GNU/Linux 12 and CUDA 12.2.
We use LLaMA-3.1-8B-instruct~\cite{grattafiori2024llama3herdmodels} as base model in the experiments.
All of our code development is based on vLLM 0.6.3, and the versions of all required packages are consistent with the requirements of vLLM 0.6.3.

\stitle{Baselines.}
We first introduce our design considerations of the baselines.
Regarding the estimator, we consider the following strategies:
\begin{itemize}
    \item \emph{W/o esitmator}: Add requests into the batch without estimating the iteration time, thereby ignoring the SLO of online requests.
    \item \emph{W/ estimator}: Add requests into the batch based on the estimation of the iteration time to meet the SLO of online requests.
\end{itemize}

Regarding the scheduler, we consider the following strategies:
\begin{itemize}
    \item \emph{FCFS scheduler}: The First-Come, First-Served strategy, which schedules offline tasks in the order they submitted.
    \item \emph{KV-cache aware scheduler}: The scheduler adopts by \ours, which schedules tasks to maximize the benefit.
\end{itemize}

Regarding the KV cache manager, we consider the following strategies:
\begin{itemize}
    \item \emph{LRU KV cache manager}: Least-Recently-Used strategy, which is the default strategy of evitor in vLLM.
    \item \emph{Task aware KV cache manager}: The KV cache manager adopts by \ours, which set the priority of KV cache according to the task type and the prefix sharing count. Moreover, it sets a threshold to control the KV cache size of running tasks to reserve space for burst requests.
\end{itemize}

We build the final \ours design by adding the components described above to the original vLLM framework. Specifically, the base implementation (BS) is the vLLM framework with priority scheduling strategy~\cite{vllm_project_pull_5958}.
We first add the estimator to BS, achieving SLO-aware scheduling (BS+E). On this basis, we extend the \ours Scheduler to be KV-cache aware and select the appropriate offline requests to conduct a batch (BS+E+S). Finally, we add the KV cache manager to implement the Task aware KV cache management strategy (BS+E+S+M, i.e., \ours).

We design different simulation methods for online and offline tasks based on their actual submission characteristics.
\stitle{Prompt datasets:}
ShareGPT~\cite{eccleston2023sharegpt} collects conversations between users and chatbots, with a low sharing rate and short prompts.
LooGLE~\cite{li2024looglelongcontextlanguagemodels} uses long articles with over 10k words as contexts and attach several questions for each long article in multiple conversations. LooGLE has a high prefix sharing rate and long prompt in batches, representing typical offline task forms.
For offline tasks, they are submitted in batched manner and we use ShareGPT and LooGLE as the datasets. We use two different subsets of LooGLE, LooGLE QA\_Short and LooGLE QA\_Long, to represent offline tasks with different length distributions.
For online tasks, we use the real-world trace and adopt ShareGPT datasets to simulate them, following the methods of existing work for online requests simulation~\cite{zhong2024distserve,agrawal2024taming,kwon2023efficient}.

\stitle{Trace:} For online tasks, we use the real-world trace of a large language model serving system, which contains request arrival timestamps. We attach sharegpt datasets to the trace to simulate the online tasks.
Note that to match the capacity of our experimental resources, we scale the timestamps of the real-world traces so that the request arrival does not cause queuing, while ensuring that the distribution characteristics of the trace remain unchanged.

\subsection{Overall Performance}
\label{subsec:collaborative_scheduling_performance}

\begin{figure}[t]
    \centering
    \includegraphics[width=0.95\columnwidth]{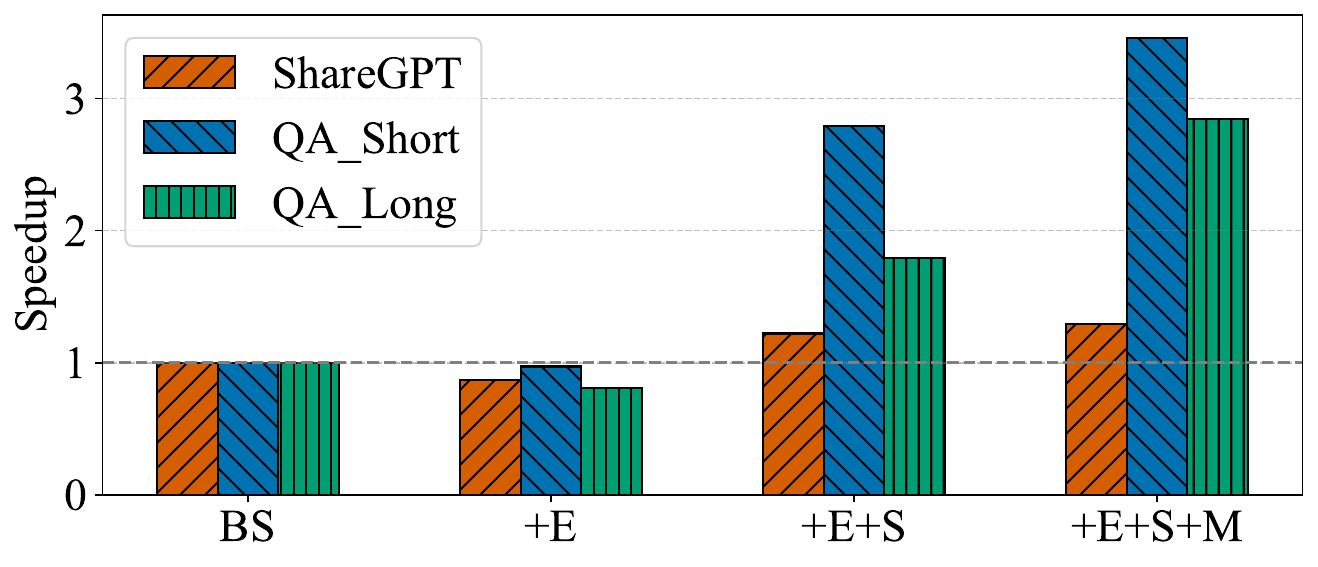}
    \caption{Throughput speedup of offline tasks compared to baseline strategies.\vspace{-.15in}}
    \label{fig:speedup}
\end{figure}

According to the different characteristics of online and offline tasks, we focus on different performance metrics.
For online tasks, we focus on TTFT and TPOT of the requests.
For offline tasks, we focus on the overall throughput that can be achieved by collaborating with online tasks.

\stitle{Throughput of offline tasks.}
We calculate the speedup of the throughput of offline tasks that each strategy can accommodate when mixed with online tasks.
As shown in Figure~\ref{fig:speedup}, the speedup of the baseline is set as 1.
BS+E has a slightly lower throughput since the SLO constraints prevent offline requests from filling the batch completely, making the GPU underutilized.
BS+E+S scheduler utilizes the prior knowledge of the length and prefix of offline tasks to select requests with the higher sharing rate with a balanced length, reducing recomputation for offline task execution and improving batch execution efficiency.
\ours further combines task aware KV cache manager to reduce the frequent flashing of the free table caused by online requests, while pinning the prefix cache with higher hit rate.
\begin{figure}[t]
    \centering
    \begin{subfigure}[b]{0.235\textwidth}
        \includegraphics[width=\textwidth]{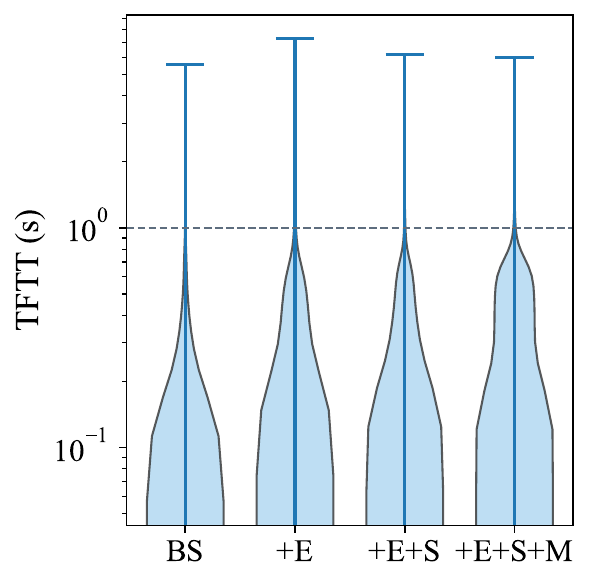}
        \caption{TTFT distributions.}
        \label{fig:violin_ttft}
    \end{subfigure}
    \hfill
    \begin{subfigure}[b]{0.235\textwidth}
        \includegraphics[width=\textwidth]{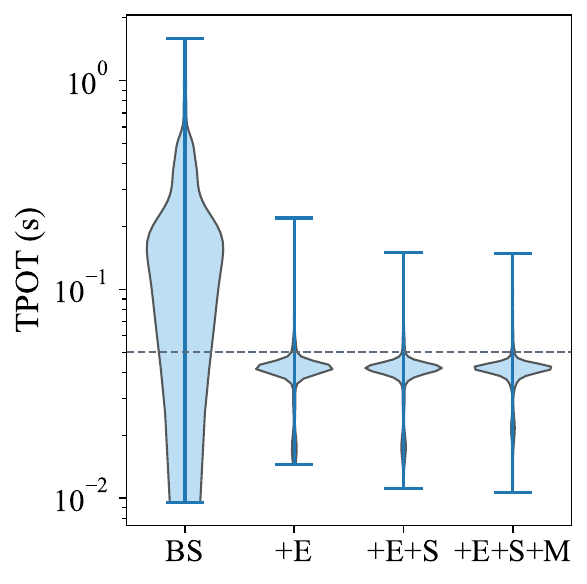}
        \caption{TPOT distributions.}
        \label{fig:violin_tpot}
    \end{subfigure}
    \caption{TTFT and TPOT distributions of online tasks.}
    \label{fig:slo}
\end{figure}

\stitle{SLOs of online tasks}
We conduct experiments on the real-world online trace~\ref{fig:trace-close} using the four strategies to verify the efficiency of our design. In the experiment, we set the TTFT threshold of online tasks to 1s, the TPOT threshold to 0.05s, and the SLO attainment threshold to 90\%. In fact, these settings will not affect the experimental conclusions since we adopt the same SLO guarantee and estimation mechanism for the estimator.
Besides, SLO is not our optimization target but a constraint.
We pre-define the scaling factor of the trace request rate through profiling and the estimator, so that it will not violate the SLO in a workload only including online tasks. Offline tasks are submitted all at once at the beginning of the experiment to provide the scheduler with enough tasks and simulate the arrival of large batches of offline tasks in actual services.
Figure~\ref{fig:slo} shows the SLO metrics of the four strategies, demostrating that all SLO-aware strategies can meet the SLO of online tasks. Note that BS achieve lower TTFTs since online requests have strictly higher priority than offline requests. The online requests in the waiting queue will be scheduled regardless of the offline requests in batch and the SLOs.

\begin{figure}[t]
    \centering
    \includegraphics[width=0.95\columnwidth]{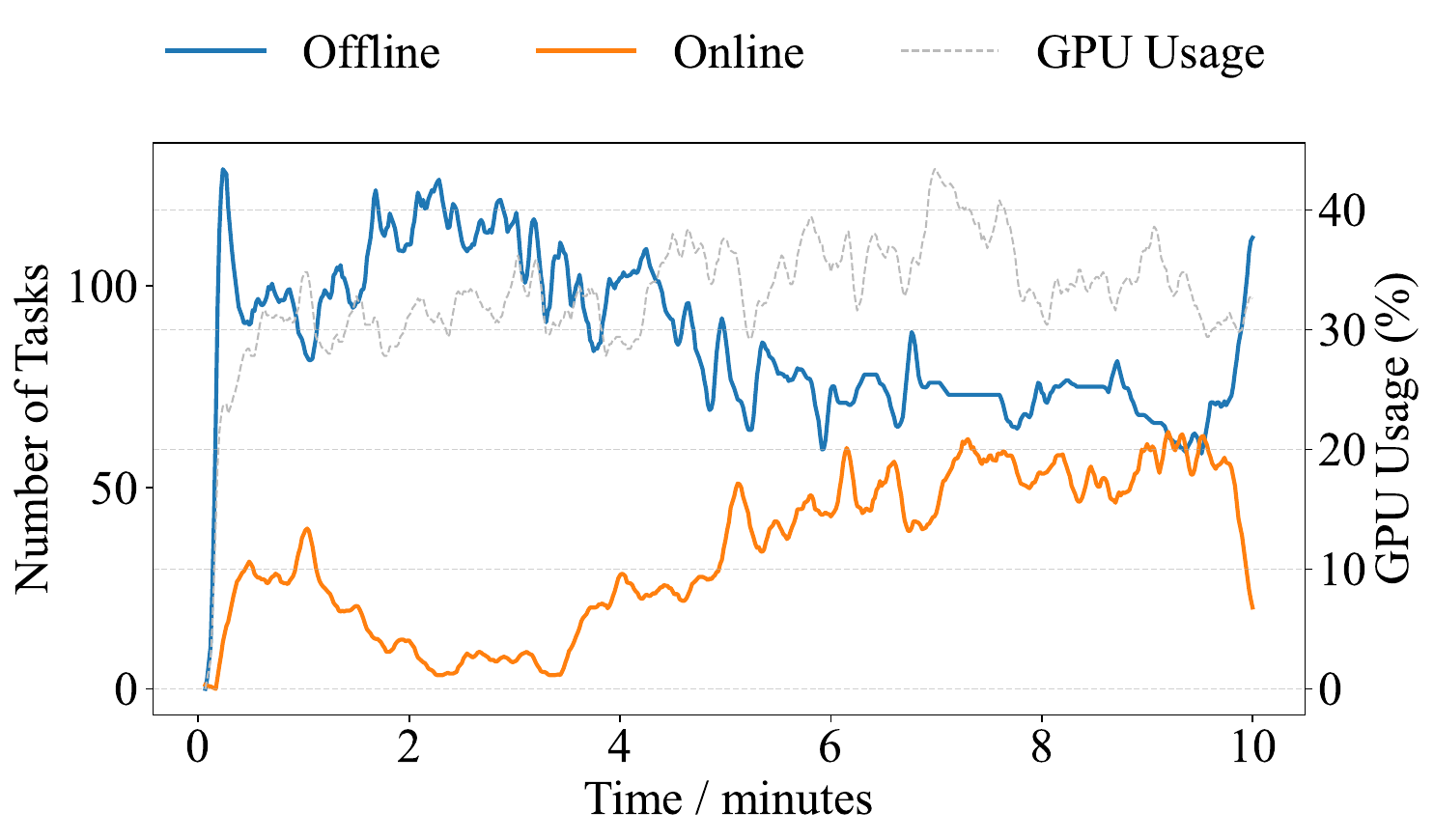}
    \caption{Throughput of online and offline tasks in the real-world trace.\vspace{-.15in}}
    \label{fig:queue}
\end{figure}

\stitle{Throughput of online and offline tasks in the real-world trace.} Figure~\ref{fig:queue} shows the active requests of online and offline tasks in the real-world trace. The online requests are bursty and have a higher priority than offline requests.
The active offline requests change in the opposite direction to the active online requests, effectively avoiding resource contention with online requests while ensuring stable GPU utilization.

\subsection{Deep Dive into KV Cache and Memory Usage}
\begin{figure}[t]
    \centering
    \includegraphics[width=0.95\columnwidth]{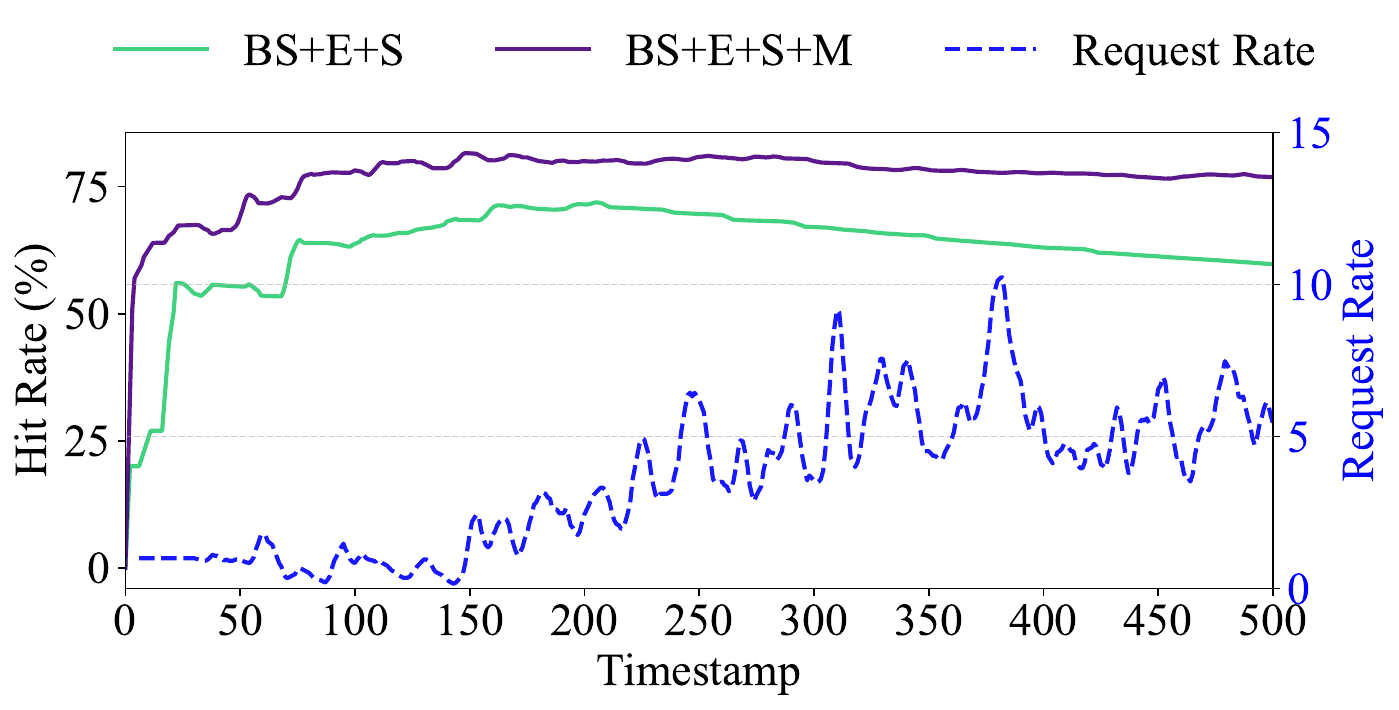}
    \caption{Prefix cache hit ratio.}
    \label{fig:hit_ratio}
\end{figure}

\stitle{KV cache hit rate.}
We compared the prefix cache hit rate of different strategies with LooGLE Short\_QA as offline workload. Since vLLM and Naive1 do not reorder offline requests, resulting in a lower prefix sharing rate, we do not include their results for fairness.
As shown in Figure~\ref{fig:hit_ratio}, from the begin of the offline tasks, \ours can achieve a higher prefix cache hit rate than Naive2, and keep the hit rate stable during the peak period of online tasks.
In naive2, the hit rate decreases as the online tasks increase, since the LRU cache eviction strategy leads to flashing of prefix cache by peak workload of online tasks, and reduces the opportunity for offline tasks to hit the prefix cache.
\ours fully utilizes the prior knowledge of offline tasks and the trace, effectively preventing the relatively inefficient online prefix cache flash and maintaining the high reuse rate of the offline prefix cache.

\begin{figure}[t]
    \centering
    \includegraphics[width=0.95\columnwidth]{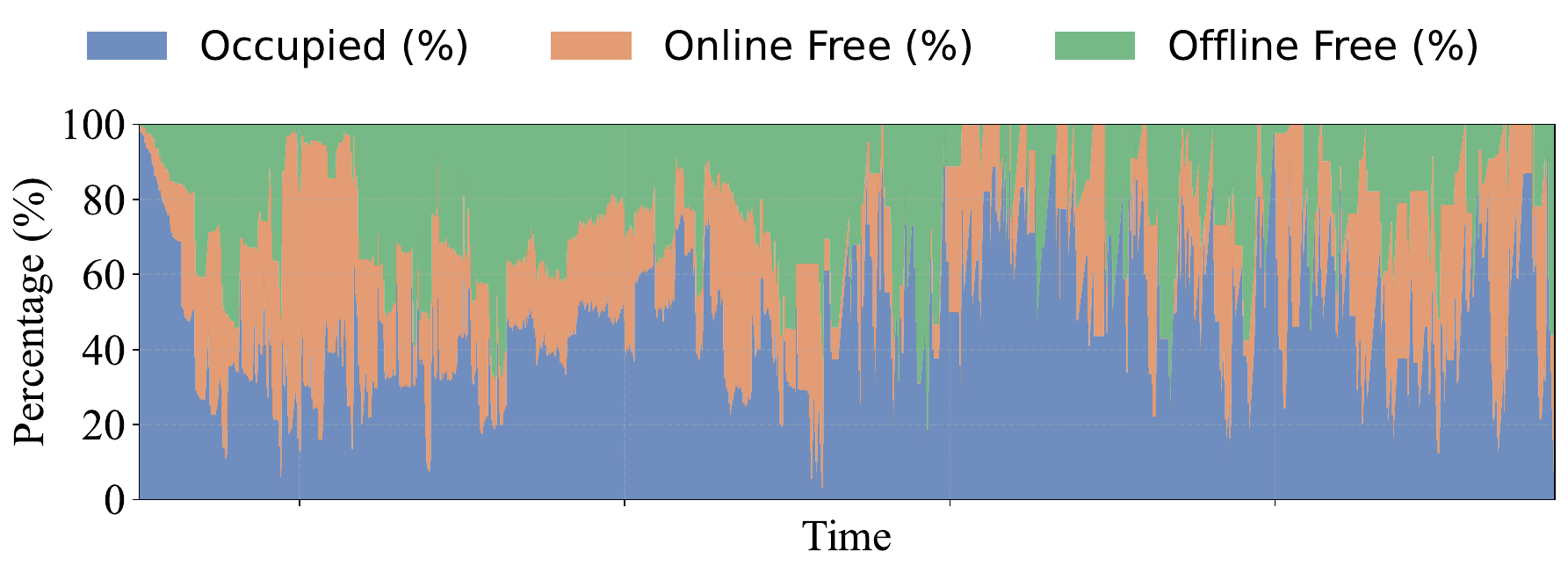}
    \caption{Memory consumption of various tasks.\vspace{-.15in}}
    \label{fig:free_table}
\end{figure}

\stitle{Memory consumption of various tasks.}
Figure~\ref{fig:free_table} shows percentage of various tasks in the memory, where the occupied memory indicates the memory occupied by the running tasks), and the online free and offline free indicate the memory of inactive online and offline tasks, respectively.
The burstiness online workload causes volatile latency, triggering the scheduler to adjust the batch size, resulting in frequent flashing of the occupied memory.
In addition, in most iterations, more than 50\% of the memory is occupied by the running tasks, indicating full utilization of the GPU memory.

\subsection{Trace Estimator}
\begin{figure}[t]
    \centering
    \includegraphics[width=0.95\columnwidth]{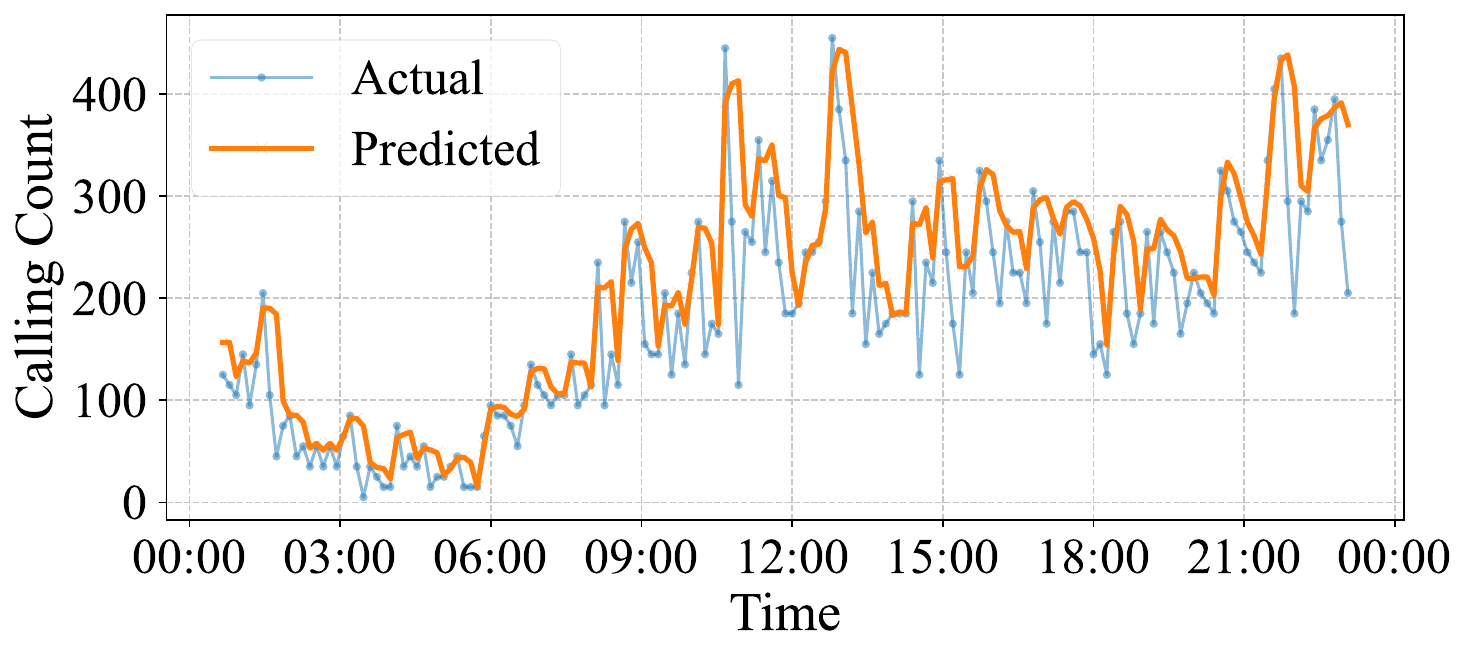}
    \caption{Predicted trace vs. actual trace.\vspace{-.15in}}
    \label{fig:Predicted_Trace}
\end{figure}

Figure~\ref{fig:Predicted_Trace} shows the predicted trace based on the real-world trace.
The estimator records request arrivals within a fixed time window (e.g. 15 minutes) and predicts the arrival rate of online requests at the current time using the standard deviation and mean of the online traces within that window. To enhance the capability to handle burst requests, we utilize a KV cache manager to reserve a certain amount of buffer space. Since bursts of requests can cause the scheduler to become SLO-bound, it is prudent to reserve additional KV cache space while providing extra space for the free table.

%% file: 8_related_works.tex
\section{Related Works}
\stitle{Task Preemption.}
Preemption is a common and effective technique to handle task with different priorities in a shared resource environment.
Conserve~\cite{qiao2024conserveharvestinggpuslowlatency} proposes a efficient preemption mechanism to reduce the resource waste of low-priority tasks by proposing a novel incremental checkpointing mechanism to mitigate the overhead of offline task swapping in and out. We notice Conserve is orthogonal to our work, and can be employed to further improve the resource utilization of offline tasks in \ours.
\cite{kim2024effectschedulingpreemptionefficiency} further studies the impact of preemption on the efficiency of LLM serving system.

\stitle{Collaborative scheduling of online and offline tasks on other cloud services.} Co-locate latency-sensitive online tasks with batched offline tasks is a widely adopted strategy to enhance resource utilization. G\"{o}del~\cite{wu2023godel} proposes to unify heterogeneous resourses across different business groups to co-schedule there online and offline workloads from various tasks, including machine learning task, data processing task, streaming task, etc.
Google's Borg~\cite{verma2015borg}, a cluster management system, permits low-priority (offline) tasks to be preempted by high-priority (online) tasks. Operating systems like Shenango~\cite{ousterhout2019shenango} also employ similar strategy to schedule threads with different characteristics.
However, none of these works focus on the emerging era of LLM serving, which has unique characteristics and challenges.

\stitle{Elastic resource allocation.} Resource allocation for online tasks is challenging due to workload burstiness and has been extensively studied in the field of cloud computing~\cite{tai2011ara,jiang2018self,jiang2020burstable,islam2015evaluating}.
The autoregressive nature of LLMs complicates resource allocation further.
SpotServe~\cite{miao2024spotserve} uses spot instances to serve LLMs, allowing elasticly allocate spot or on-demand instances. However, the allocation and initiallization of new instances is time consuming.
LoongServe~\cite{wu2024loongserve} can elastically change parallelism strategy and resource allocation for prefill and decode stage with various workloads.
As discussed in~\cite{lin2024infinitellmefficientllmservice}, if the workload is beyond the capacity, the entire task must be transferred to a instance with more GPUs, which is expensive because of KV cache mitigration; or just allocate more GPUs to the instances in the beginning, which the resources is wasted during off-preak workload.
DeepSeek \cite{deepseekai2025deepseekr1incentivizingreasoningcapability} adjusts resource allocation according to the periodic changes in inference workload between day and night. During the day, all nodes are utilized to deploy inference services, while at night, when the load is lower, the number of deployed nodes is reduced, effectively cutting costs.
In constract, \ours can fully utilize the over-provisioned resources during off-peak period by scheduling offline tasks. Moreover, \ours can further integrate with the elastic resource allocation strategy when online tasks is much more irregular and offline tasks is not enough to fully utilize the resources.

\stitle{KV cache management strategies.}
Pagedattention~\cite{kwon2023efficient} observes the fragment in GPU memory and proposes a novel KV cache memory management strategy to unlock more space for larger batch size.
Recently, researchers also consider reducing the recomputation of the same prefix of different requests by keeping and sharing the KV cache.
Preble~\cite{srivatsa2024prebleefficientdistributedprompt} studies the workload and proposes a distributed prefix caching mechanism to reduce the recomputation of different requests.
BlendServe~\cite{zhao2024blendserveoptimizingofflineinference} focuses on the offline inference and proposes a novel strategy to reduce the recomputation of same prefix of different requests by sharing the KV cache.

\section{Conclusion}

In this paper, we propose a large language model serving system, \ours, that schedules online and offline tasks collaboratively to maximize the throughput of offline tasks while ensuring the SLOs of online tasks. With the cooperation of a task scheduler, a KV cache manager, and estimation toolkits, \ours can fully exploit
1) the idle resources during off-peak workload to process offline tasks, 2) the KV cache reuse opportunities to reduce the redundant computation, and 3) batch regularity to improve the execution efficiency.
The results on real-world workloads demonstrate that \ours achieves up to $3.3\times$ throughput improvement for offline tasks compared to the baselines, while ensuring the SLOs of online tasks.